**Sharp Tunneling Resonance from the Vibrations of an Electronic Wigner Crystal**

**Authors:**

Joonho Jang[1*], Benjamin M. Hunt[1†], Loren N. Pfeiffer[2], Kenneth W. West[2], Raymond C. Ashoori[1*]

**Affiliations:**

[1]Department of Physics, Massachusetts Institute of Technology, Cambridge, MA 02139, USA

[2]Department of Electrical Engineering, Princeton University, Princeton, NJ 08544, USA

* Correspondence to: ashoori@mit.edu (R. C. A.), jjang7@mit.edu (J. J.)

† Current address: Department of Physics, Carnegie Mellon University, Pittsburgh, PA 15213, USA

Photoemission and tunneling spectroscopies measure the energies at which single electrons can be added to or removed from an electronic system[1]. Features observed in such spectra have revealed electrons coupling to vibrational modes of ions both in solids[2] and in individual molecules[3]. Here we report the discovery of a sharp resonance in the tunneling spectrum of a 2D electron system. Its behavior suggests that it originates from vibrational modes, not involving ionic motion, but instead arising from vibrations of spatial ordering of the electrons themselves. In a two-dimensional electronic system at very low temperatures and high magnetic fields, electrons can either condense into a variety of quantum Hall phases or arrange themselves into a highly ordered "Wigner" crystal lattice[4–6]. Such spatially ordered phases of electrons are often electrically insulating and delicate and have proven very challenging to probe with conventional methods. Using a unique pulsed tunneling method capable of probing electron tunneling into insulating phases, we observe a sharp peak with dependencies on energy and other parameters that fit to models for vibrations of a Wigner crystal[7,8]. The remarkable sharpness of the structure presents strong evidence of the existence of a Wigner crystal with long correlation length.

Theory suggests that a Wigner crystal (WC) may exist near integer quantum Hall states as an insulating phase with expected transition temperature in the range of a few hundred millikelvin or below[5,6,9–12]. Measurements in these regimes have demonstrated insulating phases[13–15], microwave pinning resonances[12,16], Knight shifts[17], and compressibility features[18] suggestive of a Wigner solid. However, it has proven difficult to demonstrate a long range ordered crystal; the observed features could arise from localization of electrons or a short-range ordered glassy phase. One method of detecting crystalline order is by means of observation of the crystal's vibrations. In prior measurements[1], strong coupling of tunneling electrons to vibrational modes of ions have given rise to "bosonic peaks" in the tunneling spectra. Observation of such structure was essential to verifying the BCS theory of superconductivity[1]. In this letter, we report the first observation of a sharp peak in tunneling spectra that displays the characteristics of phonons of an electron crystal.

As with a lattice formed by ions in normal materials, the WC has gapless phonon modes from transverse and longitudinal phonons. However, application of a magnetic field forces hybridizes the two modes into magnetoplasmon and magnetophonon modes[7,19], with the magnetophonon mode remaining as a gapless Goldstone mode that signals the emergence of crystalline order (spontaneous translational symmetry breaking)[7,8]. A measurement revealing the sharp magnetophonon mode would constitute direct and conclusive evidence of the existence of the WC. In this study, we detect the presence of the magnetophonon spectrum through its modification of the tunneling density of states (TDOS), and we control the energy scale of the magnetophonon spectrum by tuning the electronic carrier density and applied magnetic field. Our measurements rely on a pulsed tunneling method, capable of precisely measuring the TDOS in both conductors and insulators[20] to measure the TDOS of a 2D-hole system under high perpendicular magnetic fields (see the caption of Fig. 1 and Methods for further details).

The TDOS of a hole-doped QW measured at $B = 8$ Tesla and $T = 25$ mK are shown in Fig. 1b. In addition to the general spectral structure reported previously[20,21], these lower temperature measurements reveal very sharp resonances around the filling factor $\nu = 1$, whose full-width is about 0.2 meV. The resonances display an anti-symmetry around the Fermi level and filling factor $\nu$ (see Fig. 1c and f); i.e. for $\nu < 1$ ($\nu > 1$) the tunneling conductance peaks (dips) for energies greater than Fermi level and dips (peaks) for energies below the Fermi level.

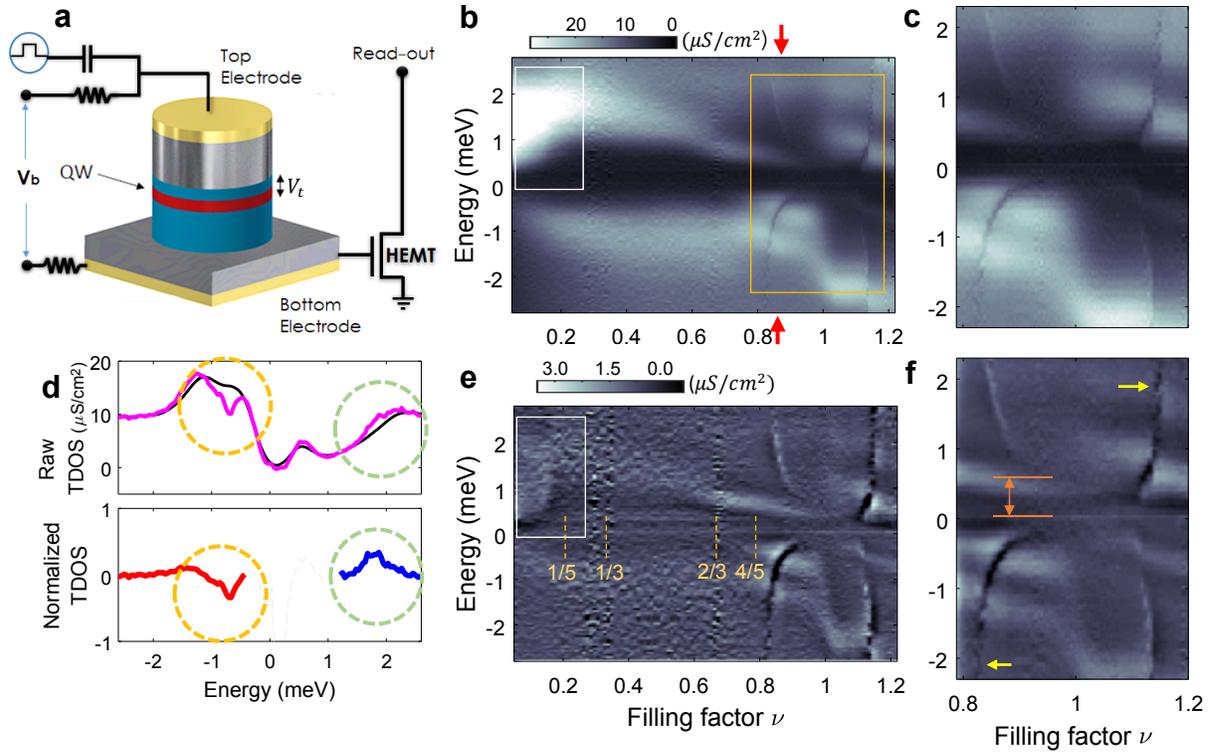

**Figure 1 | Pulsed tunneling measurement setup and TDOS of 2D holes. a,** Schematics of a vertical tunneling device. The sample is a 150 μm diameter isolated mesa made of a GaAs/AlGaAs heterostructure 14 nm wide quantum well (QW) sandwiched between two heavily doped metallic electrodes by a tunneling and an insulating barrier. A strongly correlated 2D electronic system in a QW (red) is probed by tunneling from the top electrode (silver). The bias voltage $V_b$ controls the density by drawing charge capacitively into the QW. The amplitude of an applied voltage pulse multiplied by the geometric lever-arm determines the tunneling voltage $V_t$ (see Methods). We determine the tunneling current $I$ by measuring the charge displacement by high frequency HEMT amplifiers. **b,** Tunneling density of states ($dI/dV_t$) as a function of tunneling energy $E_t = (-e)V_t$ and carrier filling factor $\nu$ (= $n_{2D}h/eB$, where $n_{2D}$ is the particle density of 2D system, $h$ is the Planck constant and $e$ is the electron charge) of a hole-doped GaAs QW at B=8 T and T=25 mK. Tunneling energies are relative to the Fermi level ($E_t = 0$) of the system, and injection (ejection) spectrum is in positive (negative) energy. The resonance in tunneling appears as the sharp u-shaped features inside the yellow box. **c,** A zoomed-in view around $\nu = 1$ is shown displaying the anti-symmetric structure of the resonances. **d,** Upper panel: Vertical line-cuts measured at 25 mK (magenta) and 250 mK (black) at $\nu = 0.86$, the location shown by red arrows in **b**. Lower panel: Normalized TDOS obtained by [TDOS(25 $mK$) − TDOS(250 $mK$)] / TDOS(250 $mK$). Anti-symmetric resonance features (dotted circles) are clearly visible. **e-f,** The same spectra as in **b-c** with a slowly-varying background subtracted to reveal fine structures of the data, respectively. The feature inside the white box shows another resonance feature near $\nu = 0$ (see also Supplementary Section 8). **f,** The orange arrow indicates the energy offset between the resonance feature in positive and negative bias (see Supplementary Section 5). The yellow arrows point to the vertical noise line discussed in the

We examine scenarios to explain this signal. First of all, the observation of the low energy resonance signals in tunneling near $\nu = 1$ may suggest an origin in spin-related excitations due to skyrmions[22,23]. However, while the strongest resonances in our spectrum appear near $\nu = 1$, we also see resonances with nearly identical energies and dispersions near $\nu \sim 0$ and $\nu \sim 2$ (See Fig. 1e and Fig. S12 of Supplementary Section 8). To our knowledge, there has been no proposal so far for spin-related excitations near filling factors 0 and 2. Secondly, we rule out disorder induced localization and oscillations of locally trapped particles due to the observed sharpness of the resonance and because these modes should soften due to screening as the quasi-particle density increases, in contrast to our observation.

It thus seems reasonable to focus on a possible electron-phonon interaction in the system. Tunneling studies show that, in metals and semiconductors, phonons can strongly couple to electrons and display resonances in electron tunneling spectrum[1]. While it is well known that tunneling electrons that *inelastically* scatter phonons can generate an increase in tunneling conductance *symmetrically* in bias voltage around the Fermi level, an *antisymmetric* feature in TDOS can arise from *elastic* coupling of electrons to the phonon degrees of freedom[1,2,24]. Formally speaking, the electron-phonon interaction generates a non-zero self-energy $\Sigma$ of tunneling electrons, whose anti-symmetric real part is strongly peaked at the energy of the van Hove singularity $\epsilon_{vH}$. The self-energy can be interpreted as an additional *complex*-valued potential only existing at an energy near $\pm\epsilon_{vH}$ that modulates the tunneling matrix element and TDOS (see Fig. 1d and Supplementary Section 1).

The emergence of these features arising from an electron-phonon interaction in the TDOS appears surprising because no strong ionic phonon DOS exists in the energy scale of a few meV -- the charged carriers in this system are effectively decoupled from ionic lattice dynamics. More importantly, the resonance energy $\epsilon_r$ in the spectrum is a strong function of filling factor $\nu$, pointing to the electron-electron interaction as the origin of the feature. These observations lead us to consider that the phonons generated from ordered states of electrons themselves affect the tunneling spectrum. In Landau levels, the existence of robust energy gaps around integer filling factors permits a description in terms of quantum Hall quasiparticles; a system with an integer filling factor $\nu = l$ (with $l = 0,1,2,...$) contains $l$ "inert" filled (spin-resolved) Landau levels, and as the filling factor $\nu$ is tuned away from integer values, the added or subtracted particles can be described as quasiparticles of effective filling factor $\nu_{qp} = |l - \nu|$. When $\nu_{qp}$ is small, these quasiparticles resemble bare electrons and holes with Coulombic interactions that may lead to the formation of a WC[6,9]

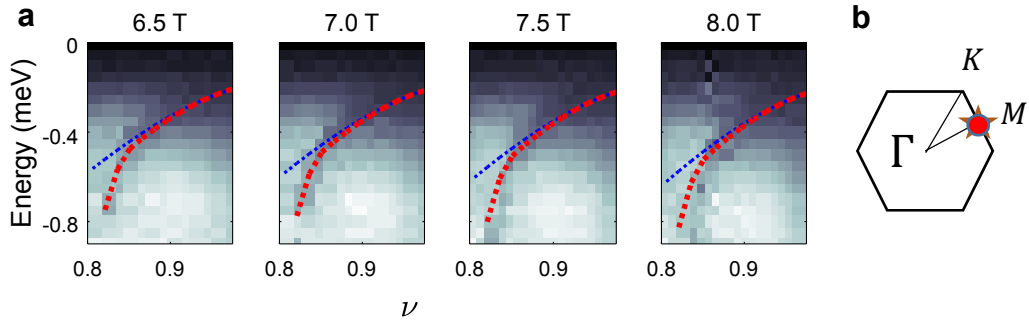

**Figure 2 | The $\nu$ dependence of the resonance energy $\epsilon_r$ at various magnetic fields. a,** The blue dashed lines are curve fits to $|1 - \nu|^{3/2} = \nu_{qp}^{3/2}$ after taking account of the Coulomb gap $\Delta_c$ (yellow dots) near Fermi level (see Supplementary Section 4 for details). The red dotted lines represent additional quantum mechanical stiffening effect (see main text for discussion). **b,** A saddle point in 2D dispersions becomes a van Hove singularity. The existence of this logarithmically-divergent van Hove singularity is a unique feature of a 2D crystalline phase. In the 2D hexagonal Brillouin zone of a WC, it is located at M-point (indicated by a red star). Note that the size of the Brillouin zone is filling factor dependent as $|\mathbf{k}_M| = \pi/l_s \propto \sqrt{n} \propto \sqrt{\nu B}$, and the overall energy scale also changes.

To examine this idea quantitatively, we explore the hypothesis that the spectrum of the resonance feature near $\nu = 1$, at low effective quasiparticle filling factor $\nu_{pq} = |1 - \nu|$, follows the energy scaling law of the magnetophonon of a magnetic-field-induced WC formed by quantum Hall quasiparticles[7,8] (See Fig. 2). In this picture, the strong resonance at $\epsilon_r$ in the TDOS results from the peaked DOS of the magnetophonons originating from a van Hove singularity (at $\epsilon_{vH}$) near the Brillouin zone boundary of the hexagonal WC. The energy of the resonance appears to follow $\epsilon_r = \Delta + \epsilon_{vH}$, where $\Delta$ accounts for the magnetic-field induced Coulomb gap $\Delta_c$ (and a spin gap in the case of injecting or ejecting minority spins - see Supplementary Section 5) near Fermi level[20,25]. Generally, in a Coulombic system we expect that the energy scale of the interaction would increase along with the effective filling factor of quasiparticles $\nu_{qp} = |1 - \nu|$ as the interparticle distance $l_s$ between the quasiparticles shrinks. In the case of a crystal, as $l_s$ decreases, the energy scale of the entire phonon dispersion and of the van Hove singularity $\epsilon_{vH}$ increase as a result. To compare our data with theoretically expected dispersions, in Fig. 2a we have overlaid our data with semiclassical and fully quantum mechanical curves calculated for phonon energy vs. filling factor. The semiclassical harmonic approximation is given by $\epsilon_{ph}(\mathbf{k}) = (e^2/4\pi\epsilon_s l_B)(\nu_{qp}\sqrt{3}/\pi)^{3/2}\sqrt{det[\widetilde{D}(\mathbf{k})]}$, where $l_B$ is the magnetic length, $\epsilon_s$ is the dielectric constant, $\widetilde{D}(\mathbf{k})$ is a dimensionless dynamic matrix whose determinant is an order of unity, and $\mathbf{k}$ is the phonon wave vector[7,8]. The van Hove singularity $\epsilon_{vH}$ is located at $\mathbf{k} = \mathbf{k}_M$ (M-point in the hexagonal Brillouin zone; see Fig. 2b) with $\sqrt{det[\widetilde{D}(\mathbf{k}_M)]} = 0.6$ [8]. This semi-classical approximation applies only when the quasiparticle density is low ($\nu_{qp} = |1 - \nu| \ll 1$). As shown in Fig. 2a, the semiclassical predictions (blue curves), after multiplication by an overall scaling factor of 1.5, fit well in the range of the filling factor $0.88 < \nu < 1$; however, as $l_s$ becomes smaller and $\nu_{qp}$ becomes larger, the quasiparticle wave functions have more overlap, and higher-order, non-harmonic, quantum mechanical effects become more important[8].

To account for this additional stiffening of the resonance, one would require the full quantum mechanical treatment. In the lowest Landau level, Chang *et. al.* have shown that crystals described with a composite fermion (CF) wave function correctly account for quantum correlations and become the lowest energy state even at very low $\nu_{qp}$[10]. For comparison with data, we have plotted curves with additional quantum stiffening expected from the shear modulus of a crystal of composite fermions with 4 flux quanta attached[11], which is shown to be the ground state below $v_{qp}$~0.18 down to $v_{qp}$~0.12 (see red curves in Fig 2; see Supplementary Section 4 for more detail). The theoretical curves capture the qualitative aspect of the stiffening of the resonance that evolves as the $v_{qp}$ increases; the nontrivial sudden increase of the stiffness in our data can be explained by the characteristic of the CF correlation near the phase transition. We also note that the short distance to the tunneling electrode (~35 nm from the well center) means that the effect of screening diminishes at higher quasiparticle densities, additionally stiffening the crystal.

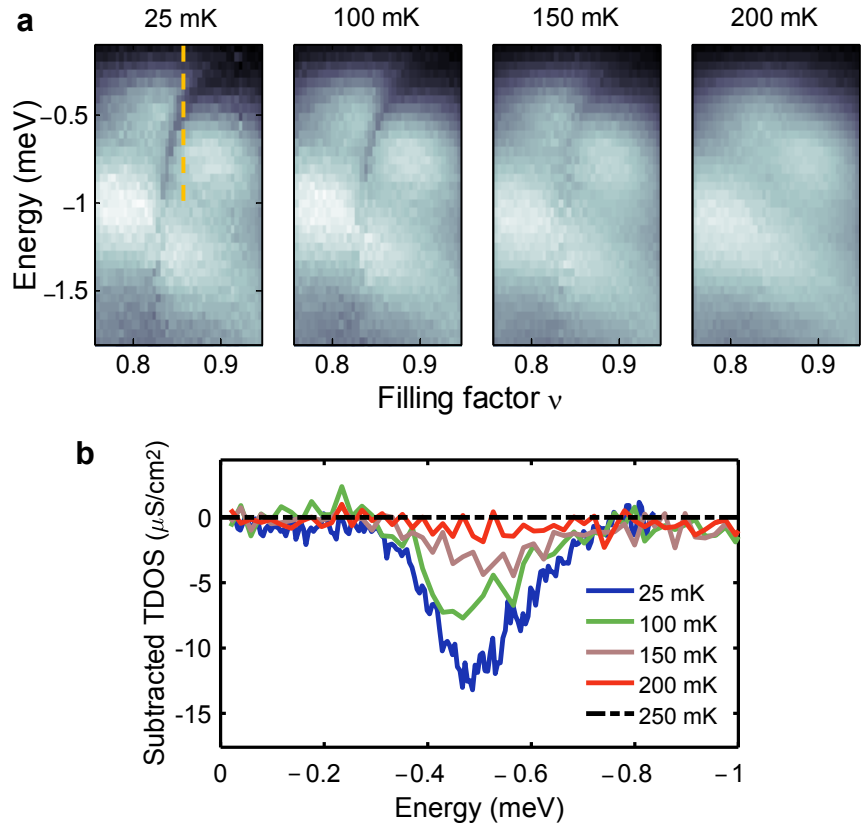

**Figure 3 | Temperature dependence of the magnetophonon resonance. a,** The resonance feature disappears above ~200 mK. **b,** Vertical line-cuts (following the yellow dotted line in a) at different temperatures after subtraction of 250 mK data as a background.

To investigate the nature of the observed resonance as further evidence of the magnetophonons and the quasiparticle crystal, we studied the temperature dependence (See Fig. 3). As temperature is increased from base electron temperature (25 mK) the resonance diminishes in strength and disappears completely above ~200 mK. This observation agrees with previous reports of the temperature induced disappearance of the insulating phases and pinning resonances and is consistent with the thermal melting of a WC of quasiparticles and the resulting disappearance of its magnetophonons[14,16]. At 25 mK, with increasing the density of quasiparticles by changing the filling factor away from $\nu = 1$, the resonance features terminate. This suggests that the WC undergoes another melting transition, one that is a purely density-driven quantum mechanical phase transition.

How the behavior of the phonon energy $\epsilon_{\nu H}$ changes as the crystal melts (either via tuning $\nu$ or $T$) can give hints on the nature of the phase transition. For a second order transition, theories predict that the shear modulus (and phonon frequency) would diminish to zero at the transition[26]; however, we observe that the magnetophonon energy $\epsilon_{\nu H}$ does not decrease toward zero in either the thermal or quantum melting transition. This strongly suggests that the phase transition is of first order with hysteresis involved at the transition[26]. Although we were not able to directly measure hysteresis involved with the first order transition, the vertical features extending from the resonances in the $dI/dV_t$ vs. $(E_t, \nu)$ plot (see yellow arrows in Fig. 1f) are consistent with a picture of increased measurement noise due to the presence of hysteresis at the transition (see Supplementary Section 7). According to a calculation of Maki et. al.[9,26] that did not account for FQH order, the shear modulus of the crystal has a peak value at $\nu_{qp} \equiv |\nu - 1| \sim$ *0.27*, then decreases to zero at $\nu_{qp} \sim 0.45$ where the crystal would undergo a second order phase transition to a liquid phase, in an apparent disagreement with our observation. However, as the authors noted, the crystalline phase competes with fractional quantum Hall (FQH) liquid states. If the ground state energy of the $\nu = 4/5$ ($\nu_{qp} = 1/5$) FQH state is lower than that of the WC, the crystalline phase would abruptly terminate before $\nu_{qp}$ reaches 0.2, with a potentially first order phase transition. Very interestingly, Archer et al. predict that the shear modulus would have discontinuous jumps at phase transitions[11] due to correlations of CFs within the crystal. We observe that the filling factor needed for the disappearance of the phonon feature is $v_{qp} > \sim 0.17$, which is in rough quantitative agreement with the theory of Archer et al. and with the termination of insulating phases previously reported in transport studies and microwave pinning measurements[11,14,16]. We are also able to identify a weaker but clear resonance signal near $v \sim 0$ in the spectrum (see the white square in Fig. 1e and Fig. S12a of Supplementary Section 8) that corroborates the fact that the WC is indeed the ground state as $v \to 0$. It is crucial to note that we measure the internal spectrum of the crystal, distinct from transport or microwave measurements of the insulating phase that indicate localization and pinning of electrons but do not demonstrate a crystalline order.

Even though we observed the resonance signals at $v = 0$ and 1, there still remains a question of whether the quasiparticles near $v = 1$ are comprised of skyrmions or normal QH quasiparticles because, in principle, phonon modes can exist in either case[27]. Skyrmions were proposed as energetically stable quasiparticles near $v = 1$ in systems where the Coulomb exchange energy scale $E_C$ dominates the Zeeman energy $E_Z$ (i.e. quantum Hall ferromagnet)[28]. Thus, some hints can be provided by measuring the electron system, where $E_Z/E_C$ is tunable by external magnetic fields, that changes the nature of the particle near $v = 1$. As an investigation into the possibility of the skyrme lattice[27], we performed similar measurements on an electron-doped sample with and without in-plane magnetic field in an attempt to tune $E_Z/E_C$. An in-plane field is expected to shrink the size of skyrmions and, at sufficiently high in-plane fields, skyrmions will no longer be the ground state quasiparticles[29]. In Fig 4, we plot spectra measured at (a) $B_\perp = 6\ T$, $B_\parallel = 0\ T$ and (b) $B_\perp = 5.8\ T$, $B_\parallel = 8.2\ T$, where $E_Z/E_C$ are 0.036 and 0.062, respectively. Thus, we expect to have small-size skyrmions for the former case and usual quasiparticles with single-spin flip for the latter[29]. We observe that, when measured without an in-plane field, the magnetophonon resonance was absent around $v = 1$. The application of in-plane magnetic fields creates visible resonances around $v = 1$, suggesting that elimination of skyrmions in favor of ordinary quasiparticles allows observation of magnetophonon resonances in the TDOS.

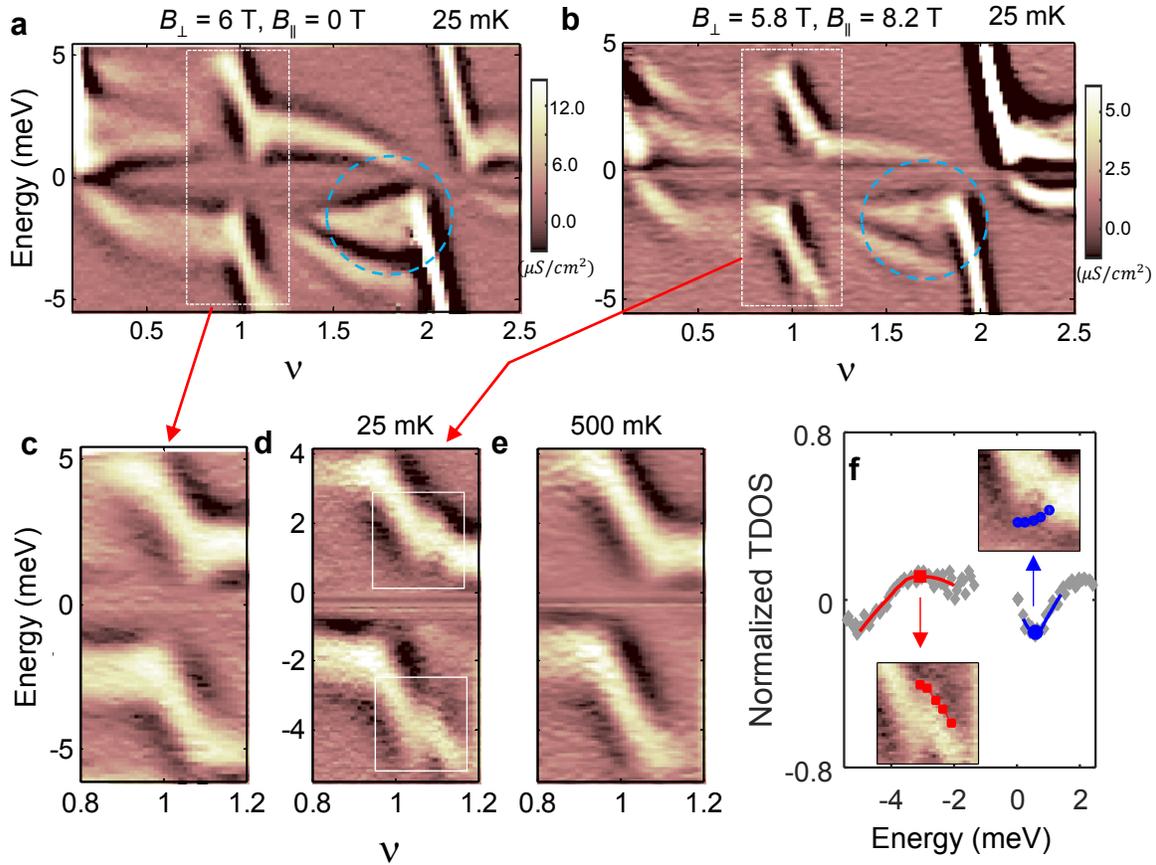

**Figure 4 | Control of the tunneling resonances of 2D electrons.** TDOS as a function of energy and $v$ (with a smooth background subtracted; see raw data in Fig. S14 of Supplementary Section 8) for a 2D electron sample. **a,** The field is 6T and applied purely perpendicularly to the 2D layer. **b,** There is a perpendicular magnetic field of 5.8 T and an in-plane magnetic field of 8.2 T. Blue dotted circles and white rectangles indicate the resonant features that exist only at low temperature. In **c** and **d**, the area inside white dotted lines in **a** and **b** are zoomed up, respectively. In **e**, temperature is increased to 500 mK otherwise same to **d**. Note the resonant features only appear at low temperature with a large in-plane magnetic field. **f,** Line-cuts of the normalized TDOS, obtained the same way as in Fig. 1**d** using 25 mK and 500 mK data at $v = 1.08$. The peaks and dips are obtained by curve-fits at various $v$'s and overlaid to the spectra inside of white rectangles in **d** with blue (dips) and red (peaks) curves as shown in insets.

Remarkably, we find that the locations of $\epsilon_{vH}$ of the electron system is very close to the ones of hole system (within 20%) while the effective mass of the electron systems is about 5 times smaller. This provides strong evidence of the characteristic mass independence of the magnetophonon spectrum in the extreme quantum limit of high magnetic fields[19] (see Supplementary Section 6).

The observed sharpness of the resonance features suggests well-defined phonon energies and hence a long lattice correlation length. For the magnetophonon mode, we expect that the energy of the resonance follows $\epsilon_{vH} \propto 1/l_s^3$. Attributing the entire width of the observed resonance to the broadening due to lattice inhomogeneity, it leads to the relationship $\Delta l_s/l_s \sim \Delta\epsilon_{vH}/3\epsilon_{vh}$. Very roughly, for the hole sample, the resonance shown in Fig. 3b gives $\Delta\epsilon_{vH}/\epsilon_{vH} \sim 0.3$, and thus $\Delta l_s/l_s \sim 0.1$, or a lattice correlation length of about 10 lattice spacings. In the Supplementary Section 3, we obtain a better lower bound estimate of lattice correlation length by numerically simulating the electron self-energy and fitting to the data. At $\nu = 0.88$, this calculation gives a lower bound of the correlation length $L_c \sim 15 l_s \sim 970\ nm$. A similar calculation for the electron sample gives $L_c \sim 3 l_s \sim 250\ nm$. We conclude by noting that this observation suggests that tunneling spectra can probe other ordered electronic states such as stripe and bubble phases in the quantum Hall regime[30].

**Acknowledgements:**

The work at MIT was funded by the BES Program of the Office of Science of the US DOE, contract no. FG02-08ER46514, and the Gordon and Betty Moore Foundation, through grant GBMF2931. The work at Princeton University was funded by the Gordon and Betty Moore Foundation through the EPiQS initiative Grant GBMF4420, and by the National Science Foundation MRSEC Grant DMR-1420541. We thank P.A. Lee and I. Sodemann for helpful conversations. We thank Neal Staley for a careful proofreading of the manuscript and Ahmet Demir for assistance in amplifier design.

**Author Contributions:**

J. J. and B. H. performed measurements, J. J. and R. C. A. analyzed data, K. W. W and L. N. P. grew GaAs/AlGaAs heterostructures, all authors discussed the results, J. J. and R. C. A. wrote the paper, and R. C. A. supervised the overall project.

**Competing financial interests**

The authors declare no competing financial interests.

**Materials & Correspondence**

Correspondence and material requests to: Raymond Ashoori (ashoori@mit.edu) or Joonho Jang (jjang7@mit.edu)

**Data Availability**

The data that support the plots within this paper and other findings of this study are available from the corresponding author upon reasonable request.

**Methods**

The measurement employs a sequence of high frequency square pulses using a Tektronix DTG5274 pulse generator and a home-made pulse shaper which minimizes distortions of the pulses (See Fig. S1a). The initial pulse induces a tunneling voltage across the tunneling barrier. Subsequently applied opposite polarity pulses are used to retrieve the tunneled electrons out of the QW.

Using a bias-tee, we are able to apply a DC voltage and pulses to the samples at the same time. The DC voltage capacitively tunes the carrier density in the 2D system (see Fig.S1a). As there is no leakage through the sample (the thick AlGaAs barrier prevents leakage), this DC voltage only changes the carrier density in the quantum well and induces no current. The filling factor (proportional to density) is calibrated precisely by independently measuring and integrating the capacitance of the samples as shown in Fig. S1d-e.

The pulse amplitude defines the tunneling energy of electrons; the tunneling energy scale is given by the pulse amplitudes multiplied by the geometric lever-arm of a heterostructure (the distance between QW to top electrode divided by the distance between top and bottom electrode), which is ~0.21 for the hole wafer, and ~0.25 for the electron wafer, respectively. The pulse parameters are adjusted to ensure the system is returns to the initial equilibrium state between pulses with predefined density of 2D system (i.e. the pulse sequence returns the system back to the same density of electrons or holes in the 2D system, with the 2D system in equilibrium with the 3D tunneling electrode - there is also a delay of approximately 300 microseconds between pulses, and this allows the 2D system back down to its base temperature). The rise-time of the pulses ~2 nsec, during which only a small number of particles (due to the slow tunneling rate in our samples - the effective RC time for equilibration is roughly on the order of 1 millisecond - about 5 orders of magnitude longer than the rise time) tunnel before the actual determination of tunneling conductance starts. This amounts to a change of $10^4$ cm$^{-2}$ in electron density. The tunneling current is determined by measuring the linear increase of the displacement charge in response to the applied initial pulse in ~2 microseconds, (during which the electron density change of ~$10^7$ cm$^{-2}$ can occur, less than a part per thousand of the equilibrium density) before the 2D system is disturbed. This means that the measurement of the tunneling current accurately measures the tunneling rate, a true equilibrium property of the sample.

The amount of charge that tunnels in and out of the QW is measured by HEMT amplifiers at the mixing chamber, and further amplified by cryogenic Si-Ge transistors (Weinreb amplifiers) and room-temperature broadband amplifiers. All amplifiers are 50 Ohm impedance-matched. The pulse sequence is repeated and averaged over 100,000~1,000,000 times for a good signal-to-noise ratio, generating a number that represents a single pixel of data in a 2D spectrum. As positive and negative energy sides are measured separately, there exists some noise near E=0. This noise doesn't affect our study and always can be effectively suppressed by measuring with finer voltage steps. To maintain low electronic temperature in the system, we thermally anchored attenuators at 1 K and 8 mK stages and used NbTi-NbTi coaxial cables between the stages. In addition, the amplifiers are thermally decoupled from the samples using a resistive manganin wire (~20 Ohm) wound around an oxygen-free high-conductivity copper heat sink post (~8 mK), and the power dissipation of the amplifier is maintained less than a few hundred nW range. By measuring the strongly temperature dependent suppression of the zero-bias tunneling in a perpendicular magnetic field, we have confirmed to achieve electronic temperature of ~25 mK ($\pm$5 mK).

# Supplementary Information (Sections #0-8)

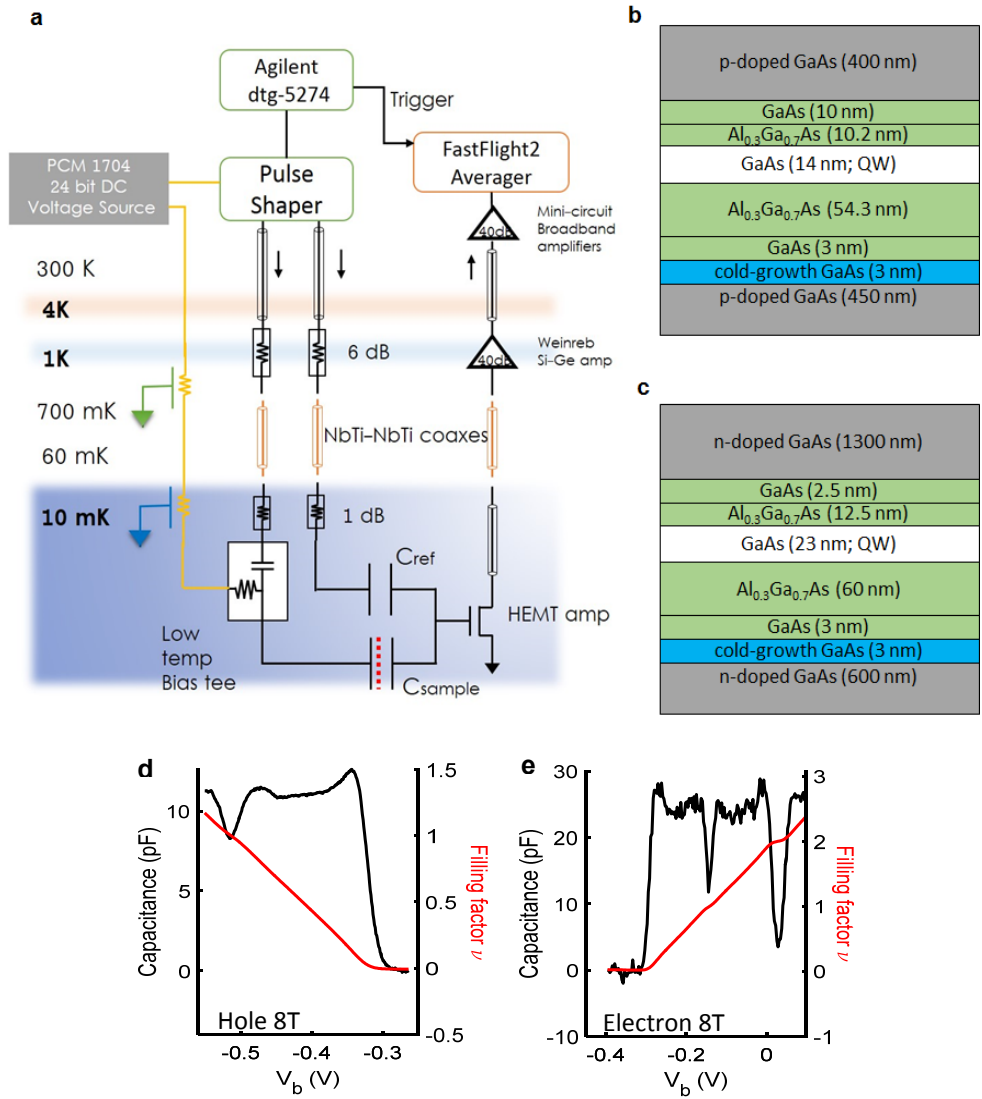

**Fig. S1 | Wafer growth sheet and measurement setup. a,** Schematics of measurement setup. **b-c,** The MBE growth profiles for wafers used in the experiment; **b,** carbon-doped 2D hole quantum well wafer (pf-5-11-12.1), and **c,** silicon-doped 2D electron (pf-11-18-13) quantum well wafers. **d-e,** Filling factor vs. bias voltage is determined by integrating the capacitance (inverse compressibility) measurements.

## 0. Samples

The samples are GaAs/AlGaAs heterostructures, and grown by molecular beam epitaxy at Princeton university. Two versions of wafer, hole-doped with Carbon (pf-5-11-12.1) and electron-doped with Silicon (pf-11-18-13), were used in these measurements. Both are grown in (100) direction. The wafers have cold-growth spacer layers (grown at 440 'C, otherwise at 620 'C; See blue colored layers in the Fig.

S1b and c) to help prevent dopant migration from the bottom electrodes into the barriers and QWs. This technique improved the quality samples over the ones used in previous studies[1] when compared by measuring the broadening of TDOS at even filling factors as shown in Fig. S2a-b. The broadening is inversely proportional to the quantum lifetime of quasiparticles in the system[2,3]. We obtain the broadening of $0.15\ meV$ for holes and $0.70\ meV$ for electrons. From these numbers, we roughly estimate the transport mobility of the hole and electron samples to be ~$8.0 \times 10^5\ cm^2/Vs$ and $1.0 \times 10^6\ cm^2/Vs$, respectively. Note that the quality of the hole sample is actually higher as the quantum lifetime is ~4 times longer. The wafers are specially designed for tunneling measurements by having a quantum well (14 nm for hole and 23 nm for electron wafer) separated by a tunneling and an insulating barrier from heavily doped electrodes. We define multiple mesas of 150 $\mu m$ in diameter using photolithography, and make electrical contacts on top and bottom electrodes for a vertical tunneling device. All measured samples (3 hole mesas from pf-5-11-12.1 and 2 electron mesas from pf-11-18-13) showed the same resonance features, respectively, within our experimental resolution.

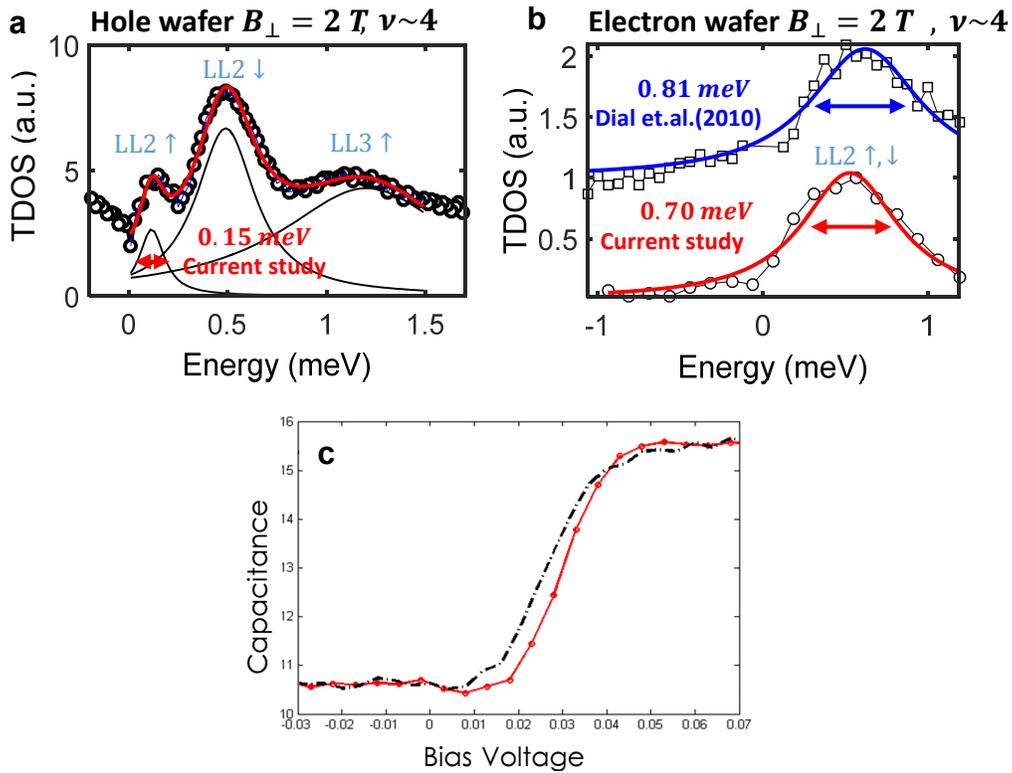

**Fig. S2 | The broadening of LL DOS as a measure of disorder strength.** **a-b,** TDOS of the hole and electron wafers measured at $\nu \sim 4$ and T=25 mK. The broadening of the TDOS near the Fermi level is proportional to the inverse of the quantum lifetime and due to disorder in the systems. **c,** The increase of capacitance of samples as hole and electrons are induced into the 2D QWs. The sharper capacitance increase means less disorder in the system. The quality of the wafers in this study (red curves) is significantly improved over the one in Dial et. al. *Nature* **464,** 566–70 (2010) (blue curve).

# 1. Calculation of tunneling matrix modification due to bosonic mode

The differential tunneling conductance $G$ (or tunneling density of states; TDOS) is written as in the zero temperature limit,

$$G = dI/dV_t = \int\int d^2\vec{k} |T(\vec{k}, E_F + eV_t)|^2 A(\vec{k}, E_F + eV_t),$$

where $T(\vec{k}, E)$ is the tunneling matrix element, and $A(k, E)$ is the spectral function of the 2D system of interest. The change in conductance can be categorized by (i) the symmetry of signal in bias voltage and

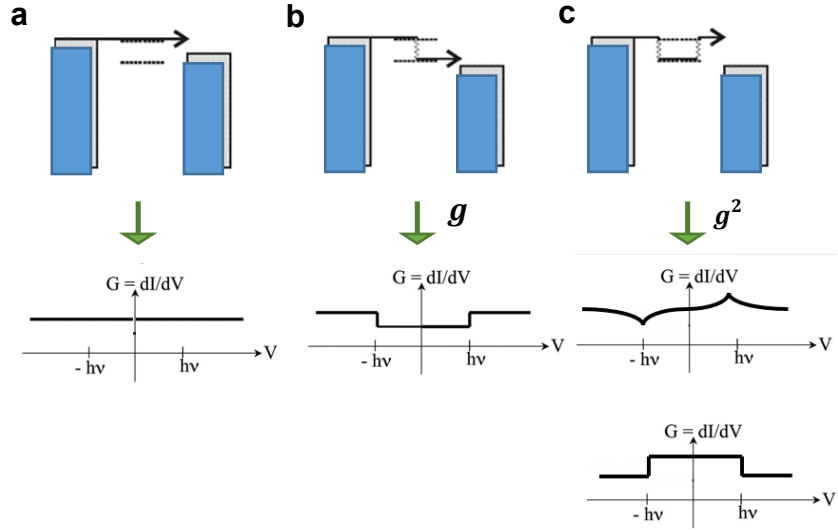

**Fig. S3 | Tunneling conductance modification due to bosonic modes (phonons). a–c**, Diagrams show the electrons interact with phonons (wavy lines) in the process of tunneling from left to right electrodes (blue) in the case of $\Delta_g = 0$. Occurrence of symmetric and anti-symmetric features in the tunneling conductance $G$ depends on the number of electron-phonon scatterings whose strength are given by $g$. Direct **(a)**, inelastic **(b)**, and elastic **(c)** tunneling mechanisms are displayed.

(ii) the existence of the energy loss in the tunneling process as follows (see also Fig. S3):

(i) The symmetry of the self-energy function dictates that the real part of $\Sigma$ generates an anti-symmetric (odd) signal in the tunneling conductance while the imaginary part produces a symmetric (even) change in conductance.

(ii) The inelastic tunneling involves the emission of real phonons, whereas the elastic conductance contribution is attributed to the emission and re-absorption of virtual phonons.

Thus, the signatures of electron-phonon (boson) coupling in the electron tunneling spectrum are a combination of peak/dip or step/anti-step features located at $\pm(\Delta_g + \epsilon_{vH})$ in $dI/dV$, where $\Delta_g$ is the single-particle energy gap near the Fermi energy. As noticed by many authors [4–7], the change in the tunneling matrix element can be large when the ratio $\Sigma/E_F$ is not small.

The self-energy term $\Sigma(E)$ modifies the tunneling matrix element because $\Sigma(E)$ acts as <u>a *complex-valued* energy-dependent potential in the 2D quantum well</u> (which we refer to as the right electrode). The argument is roughly based on the work of Taylor et. al.[6] (For more complete calculations, see Davis et. al.[4] and Appelbaum et. al.[5]). We give a simple estimate of <u>the change of the tunneling matrix $T(\vec{k}, E)$</u>, taking account of the dimensionality of electrodes and experimental parameters.

Assuming the effective masses $m_e$ of two electrodes are the same, the momentum and energy conservation law of 3D (left)-2D (right) planar tunneling leads to the following selection rules:

$$k_{l\parallel} = k_{r\parallel},$$

$$\frac{\hbar^2}{2m_e}(k_{l\parallel}^2 + k_{l\perp}^2) + E_{l0} = \frac{\hbar^2}{2m_e}k_{r\parallel}^2 + E_{r0} + \Sigma.$$

These lead to $\frac{\hbar^2}{2m_e}k_{l\perp}^2 = \Delta E_0 + \Sigma$. Here, $k_l$ ($k_r$) is the wave vector in the left (right) electrode, and the perpendicular (parallel) components to the barrier interface are indicated with suffix $\perp (\parallel)$. Also, $E_{l0}$ ($E_{r0}$) is the conduction band edge on the left (right) electrode, and the conduction band edge difference is $\Delta E_0 = E_{r0} - E_{l0}$. The non-zero self-energy $\Sigma$ in the right electrode modifies the selection of the perpendicular wave vectors $k_{l\perp}$ of the left electrode as follows,

$$k_{l\perp} = \frac{1}{\hbar^2}\sqrt{2m_e(\Delta E_0 + \Sigma)}, \quad k_{l\perp 0} = \frac{1}{\hbar^2}\sqrt{2m_e \Delta E_0}, \quad \frac{k_{l\perp}}{k_{l\perp 0}} \approx 1 + \frac{1}{2}\frac{\Sigma}{\Delta E_0},$$

where $k_{l\perp 0}$ is the wave vectors on the left electrode with the self-energy term being zero. We insert $k_{l\perp}$ into the expression of the tunneling matrix element for a rectangular tunneling barrier[7]:

$$|T(\vec{k}, E)|^2 = \frac{\kappa^2 Re(k_{l\perp}) Re(k_{r\perp}) e^{-2\kappa d}}{|\kappa - ik_{l\perp}|^2 |\kappa - ik_{r\perp}|^2},$$

where the barrier thickness is $d$, the tunneling exponent is $\kappa = \frac{1}{\hbar}\sqrt{2m(U-E)}$, and $U$ is the tunneling barrier height. Also, note that in a quantum well $k_{r\perp} = \sqrt{2mE_{c0}}/\hbar$, with $E_{c0}$ being the lowest confinement energy of the right electrode. Then, we can approximate the tunneling matrix change as

$$\frac{|T(\vec{k}, E + \Sigma)|^2}{|T(\vec{k}, E)|^2} \approx 1 - \frac{1}{2}\frac{\kappa^2 - k_{l\perp 0}^2}{\kappa^2 + k_{l\perp 0}^2}\frac{Re\,\Sigma}{\Delta E_0} + \frac{\kappa k_{l\perp 0}}{\kappa^2 + k_{l\perp 0}^2}\frac{Im\,\Sigma}{\Delta E_0}.$$

Note that $\Delta E_0$ is merely the difference of the Fermi energies between left and right electrodes, $\Delta E_0 = (E_F - E_{l0}) - (E_F - E_{r0}) = (E_{lF0} - E_{rF0}) = \Delta E_{F0}$, where $E_F$ is the Fermi level, and $E_{F0}$ is the Fermi energy defined as the energy difference between the highest and the lowest occupied states.

As a result, the changes of differential conductance $G$, normalized to the conductance without electron-phonon interaction $G_0$, are expressed as including the inelastic contribution [5,6],

$$\frac{G_{inelastic}}{G_0} = -k_F l_i \frac{\mathrm{Im}\Sigma}{\Delta E_{F0}} \qquad (1)$$

$$\frac{G_{elastic}}{G_0} = -\frac{1}{2}\frac{\kappa^2 - k_{l\perp 0}^2}{\kappa^2 + k_{l\perp 0}^2}\frac{\mathrm{Re}\Sigma}{\Delta E_{F0}} + \frac{\kappa k_{l\perp 0}}{\kappa^2 + k_{l\perp 0}^2}\frac{\mathrm{Im}\Sigma}{\Delta E_{F0}}, \qquad (2)$$

where $l_i$ is the phenomenological effective inelastic scattering length[6]. Note that the inelastic contribution is symmetric in tunneling energy, and the elastic part is a combination of anti-symmetric and symmetric component (Fig. S3).

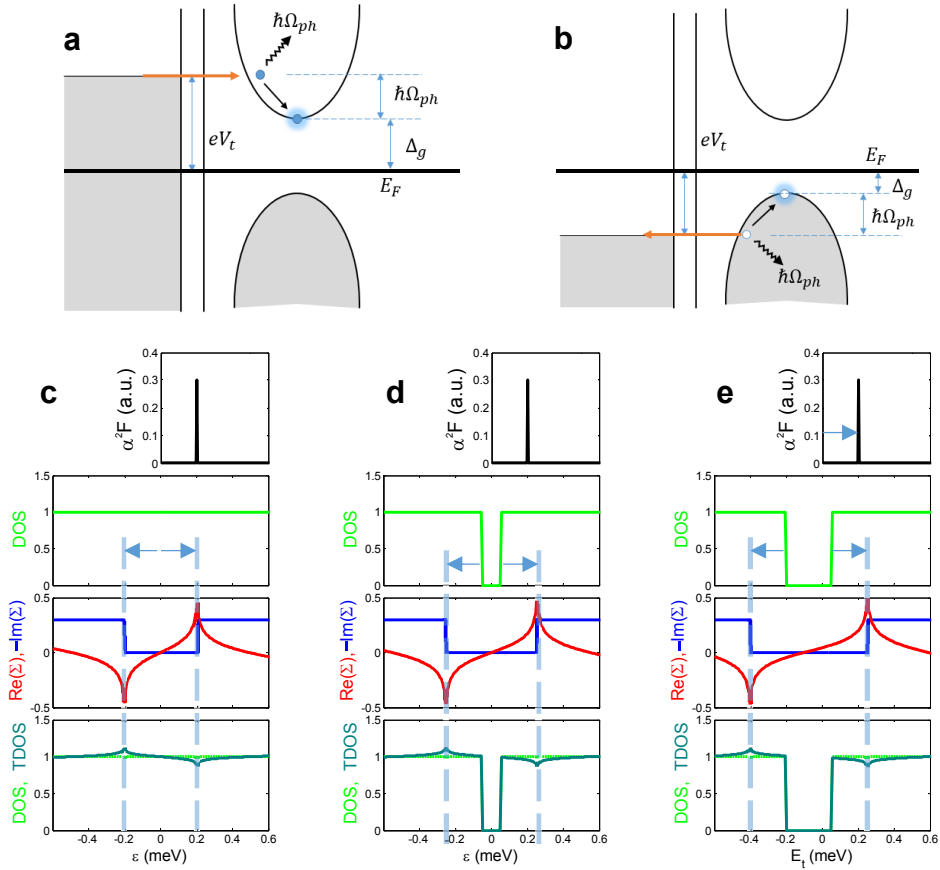

**Fig. S4 | Tunneling conductance modification due to bosonic modes (phonons).** **a-b,** Cartoon pictures that represent a tunneling electron interacting with a phonon of energy $\hbar\Omega_{ph}$ (wavy arrows) at positive **(a)** and at negative bias **(b)**. The right electrode visualizes an energy gap $\Delta_g$ near $E_F$ in the single-particle density of states of the 2D electron system under magnetic fields. When a real phonon is emitted, the tunneling process is inelastic. When a virtual phonon is emitted and reabsorbed, the tunneling is elastic. In either case, the influence of the electron-phonon interaction to tunneling signal is most pronounced when $eV_t = \pm(\Delta_g + \hbar\Omega_{ph})$, because the states above (below) this bias energy at positive (negative) bias have a probability to emit a phonon and decay into the bottom (top) of the available states. **c-e,** Simulations of TDOS with e-ph coupling with $\alpha^2 F$ of an Einstein phonon mode of $\hbar\Omega_{ph}$ and constant $N(\omega)$. No gap **(c)**, symmetric gap **(d)** and asymmetric gap **(e)** are simulated. Note the resonant features are located at $\Delta_g + \hbar\Omega_{ph}$.

In our experiment, the shape of the observed tunneling conductance strongly suggests that the elastic tunneling process is dominant over the inelastic one. We can understand this as follows. The elastic tunneling involves the two successive scatterings (emission and reabsorption) with a virtual-phonon while the inelastic tunneling process induces a single real-phonon scattering (emission). Compared to the elastic process coupled with a short-lived virtual phonon, the inelastic tunneling leads to the generation of a real phonon ($k_{vH} \sim 0.01\ \text{Å}^{-1}$) that requires a more strict condition for the conservation of energy and momentum, and is often strongly suppressed in high quality planar tunneling junctions. Thus, one needs the phenomenological effective inelastic scattering length $l_i$ in Eq. (1) to take this into account. In our devices, the planar momentum is well conserved within $0.001\ \text{Å}^{-1}$, leading to the suppression of inelastic processes. The fact that the phonon is located inside one of the electrodes, not at the barrier, additionally disfavors the traditional inelastic processes[6,8].

Then, the dominant tunneling process is basically the self-energy effect[4,7]. Based on Eq. (2) and $k_{l\perp 0} \ll \kappa$ from our experimental parameters, the dominant effect comes from the real part of the self-energy ($Re\ \Sigma$) that generates an odd (anti-symmetric) logarithm singularity feature in the $dI/dV$ spectrum. Thus,

$$G = \left(1 - \frac{1}{2}\frac{\kappa^2 - k_{l\perp 0}^2}{\kappa^2 + k_{l\perp 0}^2}\frac{Re\Sigma}{\Delta E_{F0}}\right)G_0, \tag{3}$$

And,

$$Re\ \Sigma = -\frac{G - G_0}{G_0}\frac{\kappa^2 + k_{l\perp 0}^2}{\kappa^2 - k_{l\perp 0}^2}2\Delta E_{F0}. \tag{4}$$

Our estimates of the Fermi energies of a hole-doped sample are $E_{lF0} \sim 3\ meV$ on the left 3D electrode and $E_{rF0} = 2.4\ meV$ on the right 2D electrode when density of QW is $n_{2D} = 2.0 \times 10^{11}\ \text{cm}^{-2}$ ($\nu \sim 1$ at 8 T). Thus, $\Delta E_{F0} = E_{lF0} - E_{rF0} \sim 0.6\ meV$, and the peak value of $Re\ \Sigma$ becomes as large as 0.4 meV which can account for the 30% change of tunneling conductance observed in our measurement.

Additionally, this model correctly predicts that the strength of the resonance features becomes bigger as the 2D QW density increases, because the value of $\Delta E_{F0} = E_{lF0} - E_{rF0}$ is getting smaller (note $E_{rF0}$ is linearly proportional to density of the QW); we indeed observe that the e-ph signal strength is stronger at $\nu = 1$ compared to $\nu = 0$ for hole samples. For more accurate estimate, however, a theoretical treatment including the effect of magnetic fields with Landau quantization may be necessary.

## 2. Numerical simulation of tunneling conductance

The self-energy $\Sigma$ is related to the Eliashberg function, $\alpha^2(\omega)F(\omega)$, where $F(\omega)$ is the phonon density of states, and $\alpha = \alpha(\omega)$ is the phonon coupling strength, and expressed as follows in the limit of T = 0,

$$\Sigma(\omega) = \frac{A}{N_0}\int_0^\infty d\omega'\ \alpha^2 F(\omega') \int_{-\infty}^\infty d\omega''\ N(\omega'')\left[\frac{\theta(\omega'')}{\omega + \omega' - \omega'' + i0^+} + \frac{\theta(-\omega'')}{\omega - \omega' - \omega'' + i0^+}\right], \tag{5}$$

where $\alpha^2 F(\omega)$ is the Eliashberg function, $N(\omega)$ is the electronic density of states, and $i0^+$ is an arbitrarily small imaginary number. In principle, we can experimentally determine $\alpha^2 F(\omega)$ by inverting this equation from $\Sigma$ obtained from Eq. (4). However, the inversion of the integral equation has long been known to be a mathematically unstable procedure. Instead, we can simulate the $\Sigma$ by modeling $N(\omega)$ and $\alpha^2 F$ with a sum of Lorentzian distributions, evaluate Eq. (5) and plug the calculated $\Sigma$ into Eq.(3), and compare with real data. For this, we call this effective inversion procedure the tunneling conductance "simulation" (See Fig. S4 and S5).

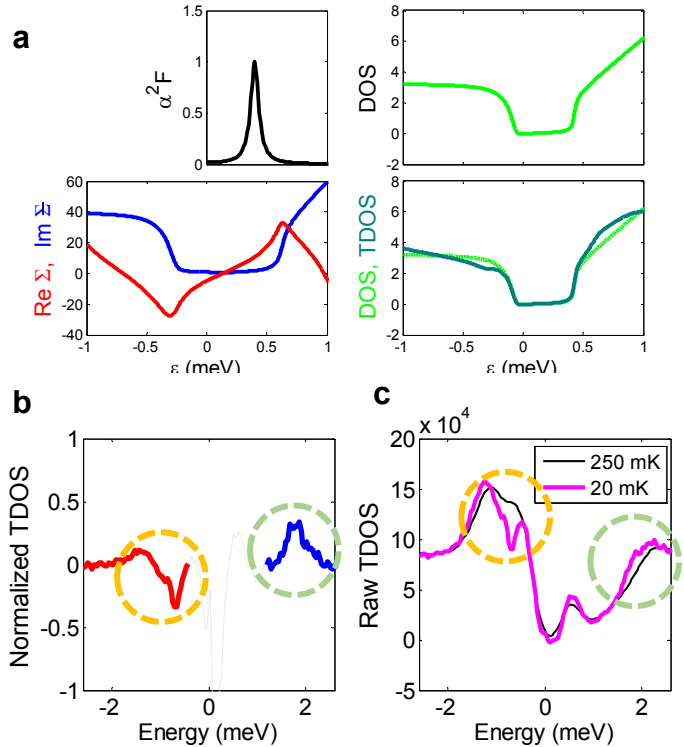

**Fig. S5 | Simulation of tunneling conductance and Eliashberg function. a,** By simulating with a broadened Eliashberg function $\alpha^2 F$ and density of states $N(\omega)$, the self-energy $\Sigma$ and the tunneling conductance are obtained. Equations of (3-5) are used. **b-c,** Reproduction of Fig.1e for comparison with the simulation. The normalized conductance ($\frac{G-G_0}{G_0}$) displays close resemblance to the simulated $Re\ \Sigma$ in **a**. Also compare the Raw TDOS in **c** to the TDOS (dark green) in **a**.

For evaluating the Eq. (5), we need to know or assume $\alpha^2 F$ and $N(\omega)$. The examples in Fig. S4**c-e** are demonstrated by assuming an Einstein phonon mode (a delta function distribution) at $\hbar\omega = \epsilon_{vH}$ for $\alpha^2 F$, and constant $N(\omega)$ with an arbitrary energy gap $\Delta_g$ near Fermi level. The most important aspect is that the resonance features are located at energies $\epsilon_r = \Delta_g + \epsilon_{vH}$. A simulation with more realistic $N(\omega)$ with broadening and an asymmetric structure in energy and $\alpha^2 F$ with a Lorentizan broadening is shown in Fig. S5, in comparison with experimental data.

## 3. Estimation of the lattice correlation length

We estimate the lattice correlation length $L_c$ by measuring the statistical broadening of the energy of van Hove singularity. From the relationship $\epsilon_{vH} \propto 1/l_s^3$, we have $\frac{\Delta\epsilon_{vH}}{\epsilon_{vH}} = \Delta\left(\frac{1}{l_s^3}\right)/\frac{1}{l_s^3}$, and $\Delta\epsilon_{vH}/\epsilon_{vH} = -3 \times \Delta l_s/\bar{l}_s$. Thus,

$$L_c \sim (\Delta l_s)^{-1} \sim \frac{\epsilon_{vH}}{\frac{1}{3}\bar{l}_s \Delta\epsilon_{vH}} \tag{6}$$

, where $\Delta\epsilon_{vH}$ is the statistical distribution of the van Hove singularity energy, $\Delta l_s$ is the statistical distribution of lattice spacings, $\bar{l}_s \sim \sqrt{2\pi}l_B/\sqrt{\nu}$ is the mean value of $l_s$, $l_B$ is the magnetic length, and $L_c$

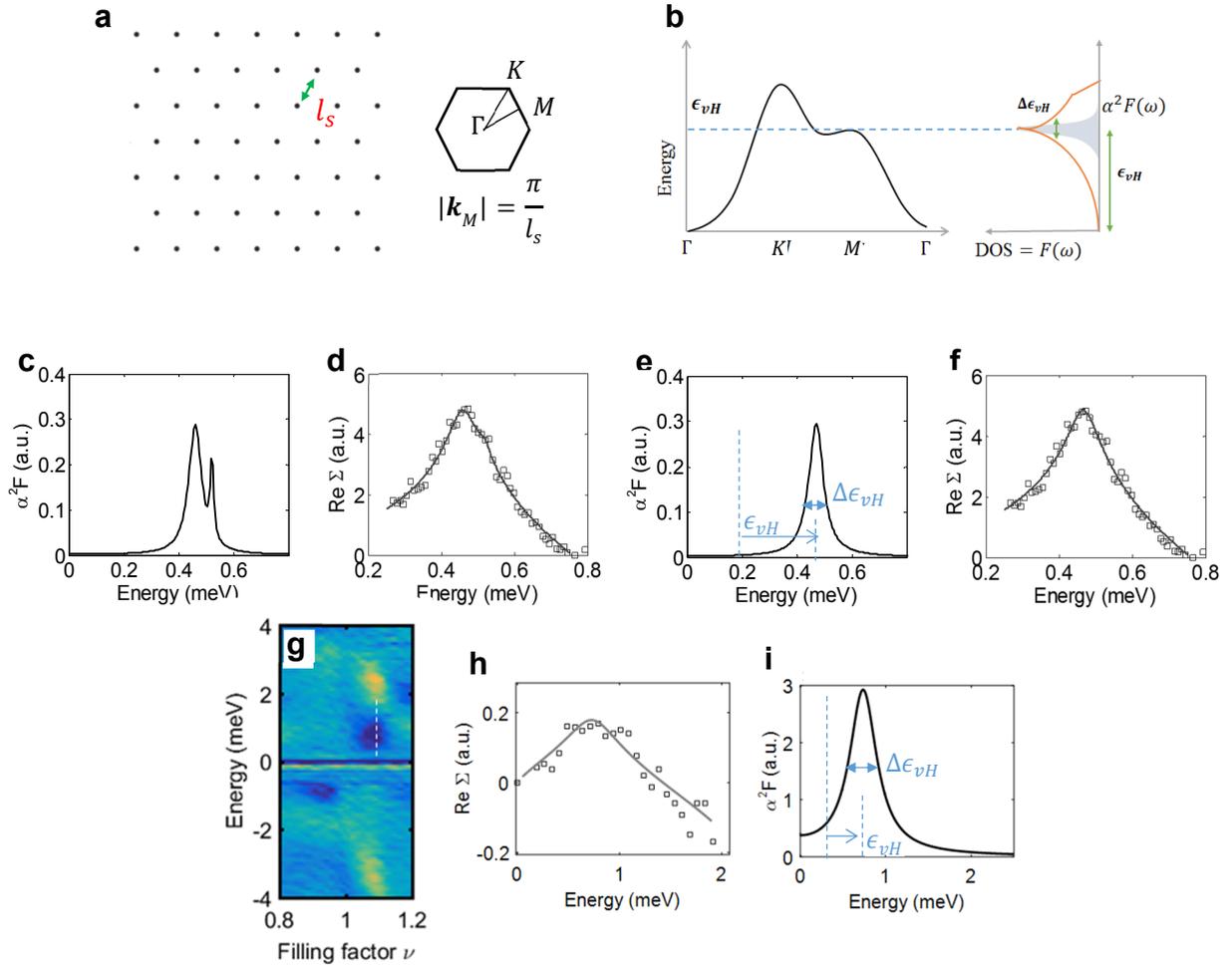

**Fig. S6 | Simulation of tunneling conductance and Eliashberg function. a,** The real space configuration of a hexagonal lattice and corresponding Brillouin zone. **b,** A cartoon representation of the possible magnetophonon dispersion, density of states and Eliashberg function. Shaded region ($\alpha^2F(\omega)$) displays the probable Eliashberg function. Note that it is not necessarily same as the phonon DOS. **c-f,** Curve fits of the real part of the simulated self-energy to a normalized TDOS data. **d,** The self-energy is calculated by modeling the Eliashberg function $\alpha^2F$ with the two Lorentzian distributions in **c**. Height, width and center location of the Lorentzian functions are fitting parameters. **f,** the curve fit with a single Lorentzian in $\alpha^2F$ in **e**. **g-i,** The same calculation for the electron system shows a shorter correlation length (see the text below).

is the lattice correlation length.

We can estimate $\epsilon_{vH}$ and $\Delta\epsilon_{vH}$ from the location and broadening of $\alpha^2 F$ which strongly peaks at $\epsilon_{vH}$. The Eliashberg function $\alpha^2 F$ can be estimated by performing a tunneling conductance simulation. We model a simple $N(\omega)$ with a sharp gap edges at $\Delta_g$, and $\alpha^2 F$ with a sum of several Lorentzian distributions. Then, with the $N(\omega)$ and $\alpha^2 F$, we can calculate $Re\,\Sigma$ using Eq. (5). This calculated $Re\,\Sigma$ is compared with measurement data for a least square regression fit process. The fitting parameters are locations, widths, and heights of Lorentzian distributions of $\alpha^2 F$ (see Fig. S6**c-f** for the results), so the number of parameters of the fit are 3 ×(number of Lorentzians). In Fig. S6, two Lorentzians are used in (f), and one is used in (h). The curve fit is slightly better with two Lorentzians, and the difference in estimation of the broadening of $\alpha^2 F$ is only ~10 percent. Thus, we chose the value obtained from the fit using two Lorentzians. Finally, we would equate the energy broadening of $\alpha^2 F$ to $\Delta\epsilon_{vH}$, and calculate the distribution of lattice parameter $\Delta l_s$ from Eq.(6).

By fitting the self-energy and obtaining the broadening of the modeled Eliashberg function $\alpha^2 F$, we get $\Delta\epsilon_{vH}/\epsilon_{vH} \sim 20\,\%$. Thus, we estimate $\Delta l_s/\bar{l}_s \sim 7\,\%$ and the crystal correlation length $L_c \sim 15\,\bar{l}_s$. Because $\bar{l}_s$ at $B_\perp = 8\,T$ and $\nu = 0.88$ (i.e. $\nu_q = 0.12$) is about 65 nm, this corresponds to $L_c = 15 \times 65\,nm \sim 970\,nm$.

We performed the same estimation for the electron system (Fig. S6**h-i**). With $B_\perp = 5.8\,T$ and $\nu = 1.1$ (i.e. $\nu_q = 0.1$), the quasiparticle lattice spacing is $\bar{l}_s \sim 85\,nm$. Also, from the Eliashberg function fit we get $\Delta\epsilon_{vH}/\epsilon_{vH} \sim 85\%$, it gives $\Delta l_s/\bar{l}_s \sim 30\,\%$. Thus, it is about 2.5~3 lattice spacing of correlation with the correlation length $L_c \sim 3 \times 85\,nm \sim 250\,nm$.

It is important to note that this estimation will give only the absolute lower bound of the lattice coherence length $L_c$; $\alpha^2 F$ has an intrinsic width in energy (even in a perfect crystal), i.e. $\Delta(\alpha^2 F)_{observed} = \Delta\epsilon_{vH} + \Delta(\alpha^2 F)_{intrinsic}$, and $\Delta\epsilon_{vH}$ is always smaller than $\Delta(\alpha^2 F)_{observed}$. Thus, by estimating the actual disorder-induced broadening $\Delta\epsilon_{vH}$ from $\Delta(\alpha^2 F)_{observed}$ from the simulation, we would get a value for $L_c$ that is smaller than the actual value.

## 4. Details of curve fits for $\epsilon_r$ on $\nu^{3/2}$

In Fig.2 and Fig. S7**a**, we show curve-fits of the measured resonance energy $\epsilon_r$ to the functional form of $\nu^{3/2}$ (referenced to the Coulomb gap $\Delta_c$) in various magnetic fields. The location of the resonance $\epsilon_r$ as a function of filling factor $\nu$ is fitted with curves of the following form:

$$\left(\frac{e^2}{4\pi\epsilon_s l_B}\right)\left\{C_1 + C_2\left(\frac{\sqrt{3}\nu}{\pi}\right)^{3/2}\right\} = \Delta_g + \epsilon_{vH}.$$

Two field-independent fit parameters $C_1$ and $C_2$ are used to fit all field data. The location of the gap edge ($\Delta_g$) is determined by locating a peak in $d^2I/dV_t^2$ spectrum, which corresponds to the sharp increase in $dI/dV_t$ spectrum. And, the location of the resonance is obtained by fitting with the local extrema in $dI/dV_t$ data.

The horizontal dotted yellow lines in Fig. S7**a** represent the magnetic-field-dependent Coulomb gap $\Delta_g = \Delta_C = (e^2/\hbar l_B)C_1$, and the red dotted curves indicate the fit curves for $\epsilon_{vH} = (e^2/4\pi\epsilon_s l_B)(\sqrt{3}\nu/\pi)^{3/2}C_2$. The fact that we have only two free parameters $C_1$ and $C_2$ throughout the all magnetic fields data indicates that the curves are functions of only $E_c = e^2/4\pi\epsilon_s l_B$ and $\nu$ regardless of magnetic fields. The fitting of the semi-classical formula (blue dashed lines) is done in the range $0.9 < \nu < 1$, where the fitting parameters are $C_1 = 0.0165$ and $C_2 = 0.093$. The deviations at $\nu < \sim 0.88$ are obvious.

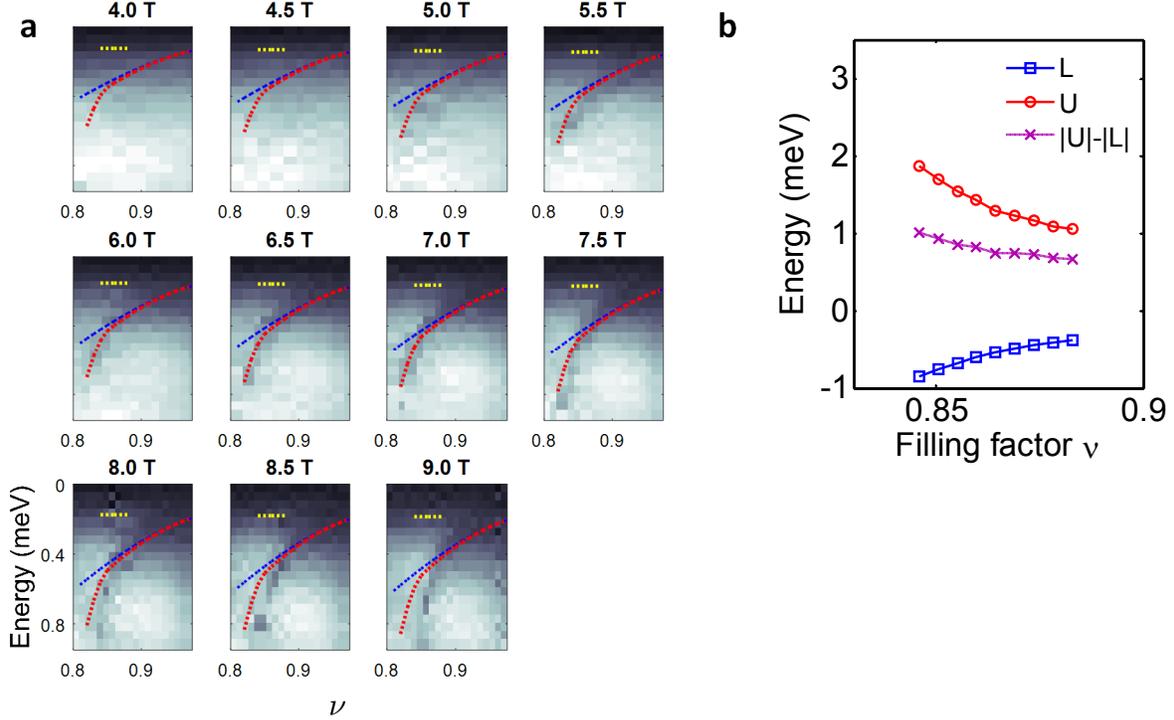

**Fig. S7 | Curve fits to resonance features. a,** A complete data set for Fig.2a of the main text. The magnetophonon energy is scaled as $\omega_- \sim 1/l_s^3 B \sim n^{3/2}/B \sim \sqrt{B}\, \nu^{3/2}$. This power law is the distinctive feature of the magnetophonon spectrum. The blue lines are fits based on the semi-classical formula and the red lines are to account for the additional stiffening due to quantum mechanical correlation calculated based on the theory of composite fermion crystal (see text). **b,** The energy locations of the resonance features in TDOS. The feature in positive (negative) energy bias is labeled as U (L), and the subtraction ($|U| - |L|$) between the locations of the two features is also shown.

The curves with an additional quantum mechanical effect included (red dotted lines) are based on the shear modulus of the composite fermionic crystals in the lowest Landau level calculated by the quantum Monte-Carlo simulation of Archer *et. al.*[9]. We estimated the energies of the phonon modes by assuming that they are proportional to the square-root of the shear modulus, $\sqrt{C_t}$; we believe this is a qualitatively reasonable assumption beyond semi-classical limit. The composite fermionic crystal with 4 flux quanta attached is the lowest energy wave function at $0.12 < \nu_{qp} < 0.18$, and still very close to the ground state even below $\nu_{qp} < 0.12$. The numbers beyond the applicable region of the simulation (i.e. $0.18 > \nu_{qp} > 1/6$) were obtained by extrapolation, which is semi-quantitatively validated from the behavior of the shear modulus in this regime from the analytic solution of the same problem[10]. The theories also show that near $\nu_{qp} = 0.18$ the crystal of 4-CF (composite fermions with 4 flux quanta attached) will go through a transition to the crystal of 2-CF (one with two flux quanta attached) before 1/5 FQHE is reached. This theoretical prediction appears to be consistent with the filling factor where the tunneling resonance feature disappears. The strong increase of stiffness of the crystal shown in the simulation can explain the resonance energy changes near $\nu_{qp} \sim 0.18$.

## 5. Interpretation of the offset energy near $\nu = 1$

The resonances display small energy offsets involved in injection (for $\nu < 1$) and ejection (for $\nu > 1$) spectra (See orange arrows in Fig. 1f). The qualitative behavior of this observed signal is consistent with the quasiparticles interacting with the magnetophonons. A naive expectation would be that the magnetophonon energy $\epsilon_{\nu H}$ goes to zero (and thus $\epsilon_r \to \Delta_C$) as the quasiparticle density vanishes approaching $\nu = 1$ (i.e. $\nu_{qp} \to 0$). However, since the scattering of the virtual (magneto)phonon is a spin-preserving process, the high energy quasiparticles can only relax into the states with same spin (in the short time scale allowed for the virtual process). Around $\nu = 1$, the exchange enhanced spin-gap exists and is expected to be the main source of the energy offset generating the offset asymmetry of the resonant features. Specifically, in the case of injecting particles with minority spins, the lowest lying available states are located higher than Fermi level by the spin exchange gap $\Delta_s$ near $\nu = 1$.

In both cases of the hole system and the electron system with a large in-plane field, our view is that the Wigner crystal is very nearly fully or fully spin-polarized for filling factors near $\nu=1$. As shown below in the diagram, there are four possible cases of injections and ejections (injection of "anti-particle"; we use this term instead of "hole" to avoid possible confusion to positively charged doped carriers), where the tunneling particle goes to/come from interstitial space; the observed resonances are consistent with these cases.

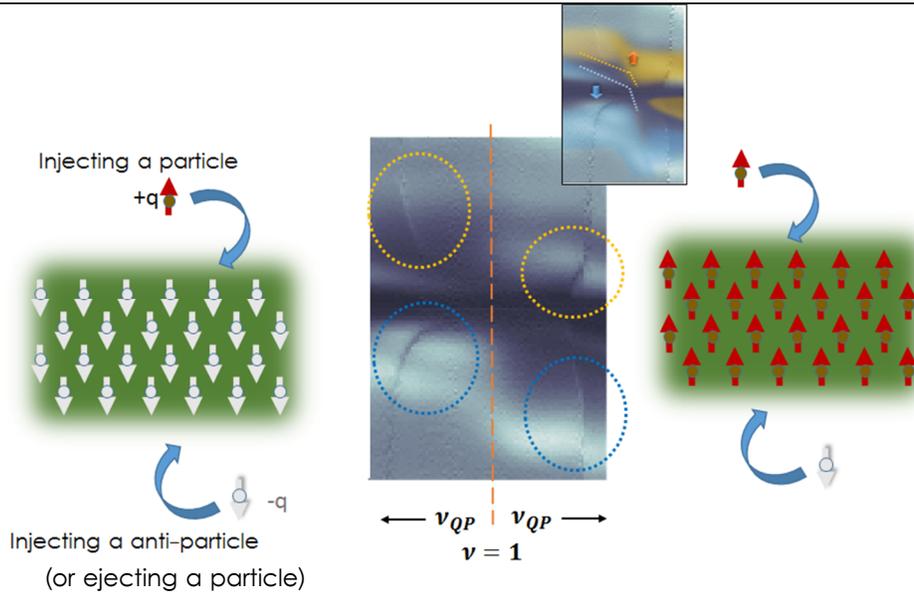

**Fig. S8 | Spin configurations near $\nu = 1$.** The offset of the resonance features suggest that injecting with spin-up particles show the resonance experiencing the additional spin energy (spin-gap), while the injecting with spin-down particles do not generate resonance possibly because the particles are force to fill the crystal lattice sites.

We take the example of $v < 1$ (we believe these arguments to apply to both $v < 1$ and $v > 1$). For $v < 1$ the bulk has spin-down particles and the vacancies that comprise the Wigner Crystal (, which we have labeled with white spin-down arrows). The Fermi level is pinned to the spin down states as these are the states as this is the spin state being filled for $v < 1$. When a particle is ejected, we then expect it to be a spin-down particle. Upon injecting a particle, there are two possibilities. There are vacancies in the spin-down background that could be filled with spin-down particles, and there are sites at higher energy that could be filled with spin-up carriers. The data show a shift-up energy for injection, indicating that the observed signal arises from injecting spin-up carriers. We believe this arises simply because there are many more spin-up sites to fill than spin-down sites (the spin-down sites for injection are the vacancies (lattice sites) that comprise the Wigner Crystal).

The measurement of the asymmetry of the locations of the features above and below Fermi level will give information about the spin gap near $v = 1$. As shown in Fig. S7**b**, the spin gap energy can be extracted as the separation of two locations ($|U| - |L|$). Surprisingly, we find that the spin-gap decreases as we approach $v = 1$. This may be a feature of hole-doped samples that likely contain substantial Landau level and subband mixing.

## 6. Magnetophonon, magnetoplasmon and mass-independence

A two-dimensional electron crystal has one transverse and one longitudinal phonon vibrational modes in the absence of the magnetic field. When a perpendicular magnetic field is applied, the two modes are hybridized into magnetophonon and magnetoplasmon because the electrons experience the Lorentz force and circulate into cyclotron orbits[11,12]. The magnetophonon is the gapless Goldstone mode while the magnetoplasmon is gapped with the cyclotron energy. In this section, we give a brief summary of the modes and show that the energy scaling law of the magnetophonon is proportional to $n^{3/2}/B$.

The displacement of the electron $r_i$ at the lattice point $R_i$ is expressed in its Fourier component $r_k$ as

$$r_i = \frac{1}{\sqrt{N}} \sum_k r_k e^{i\hbar k \cdot R_i}.$$

In an oscillatory motion $r_k(t) = r_k e^{i\omega_q t}$, the equation of motion in the absence of magnetic field is given as,

$$-\omega_k^2 r_k + \hat{C}(k) r_k = 0.$$

Here, $\hat{C}(k)$ is the dynamical matrix which has the information about the Coulombic restoring force of the system. After solving the equation by diagonalizing the dynamical matrix $\hat{C}(k)$ [12], we can express the equation of motion with a new diagonal matrix $\hat{D}(k)$ in new basis vectors $u_k$,

$$-\omega_k^2 u_k + \hat{D}(k) u_k = 0,$$

$$\hat{D} = \begin{pmatrix} \omega_{Tk}^2 & 0 \\ 0 & \omega_{Lk}^2 \end{pmatrix}.$$

Here, $\omega_{Tk}$ ($\omega_{Lk}$) is the transverse (longitudinal) mode frequency. The full solutions are given by Bonsall et. al.[12], and the overall energy scale of the modes are proportional to $\omega_0 = (8e^2/l_s^3 m)^{1/2}$.

Our interest is the case with a perpendicular magnetic field. In a magnetic field, the Lorentz force couples the two otherwise independent modes, and the new equation of motion is[13]

$$-\omega_k^2 u_k + i\omega_k \omega_c u_k \times \hat{z} + \hat{D}(k) u_k = 0 \,. \tag{7}$$

The non-trivial solutions of the equation of motion should satisfy

$$\begin{vmatrix} \omega_{Tk}^2 - \omega_k^2 & i\omega_c \omega_k \\ -i\omega_c \omega_k & \omega_{Lk}^2 - \omega_k^2 \end{vmatrix} = 0 \,.$$

The two solutions $\omega_-$ and $\omega_+$ are called magnetophonon and magnetoplasmon, respectively, and $\omega_c$ is the cyclotron frequency. In the limit of $\omega_c \gg \omega_T, \omega_L$,

$$\omega_-(k) = \frac{\omega_T \omega_L}{\omega_c} = (\frac{e^2}{\hbar l_B})(\frac{\nu\sqrt{3}}{\pi})^{\frac{3}{2}} \sqrt{\det[\tilde{D}(k)]} \,,$$

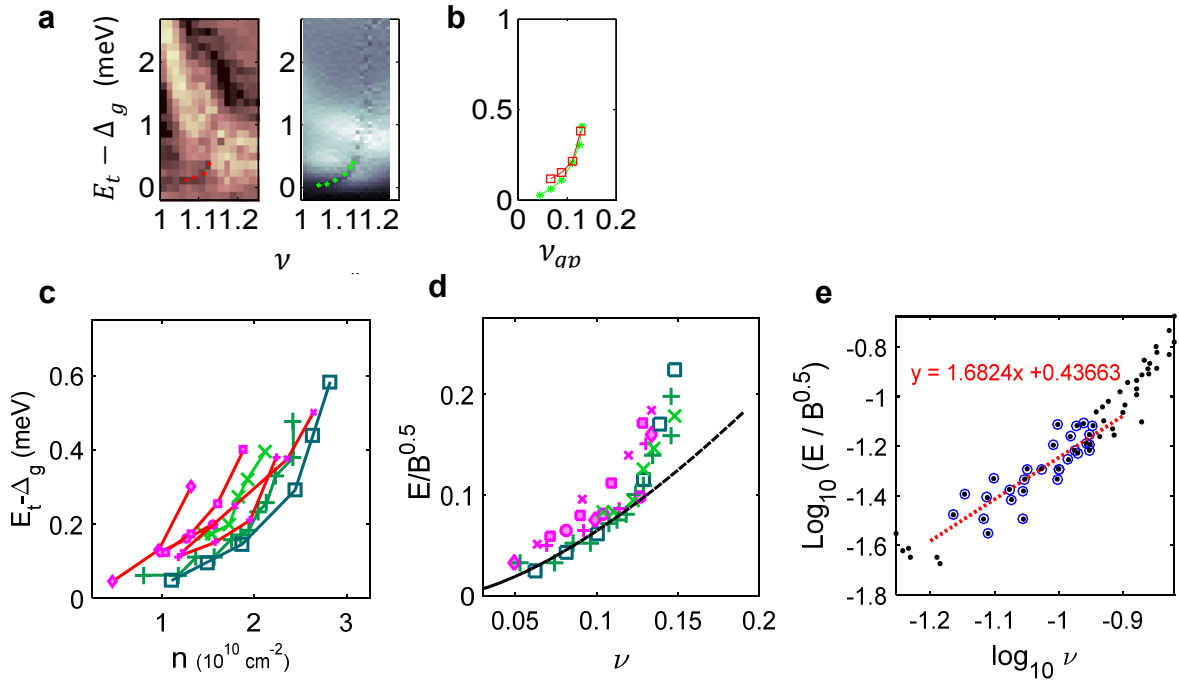

**Fig. S9 | Log-Log plot of the power law $\nu^{3/2}$ and mass independence of phonon dispersion. a-b,** A comparison of the energy scale of electron and hole samples at 7 T. The energy referenced to the gap $\Delta_g$ is displayed as a function of filling factor. The dotted lines are guides to the eye and separately plotted in **b** for a direct comparison. **c,** The locations $\epsilon_{vH}$ of the resonance features of electrons ($B_\perp$ = 5, 6, 7, 8 Tesla; magenta markers) with in-plane fields and those of holes ($B_\perp$ = 6, 7, 8 Tesla; green markers) are plotted as a function of density. **d,** $E_t - \Delta_g$ is divided by $\sqrt{B}$ as a Coulomb energy scale and plotted against $\nu_{qp}$. Note all data points collapse onto the predicted $\nu^{3/2}$ curve. **e,** The same data are displayed as a log-log plot. After taking account of the 17% difference in magnitude of electron and hole data (a shift of -0.08 of electron data in the log plot), the least square fit of the data in $0.05 < \nu_{qp} < 0.11$ (indicated by blue circles) shows the power exponent of $1.68 \pm 0.081$, which demonstrates a reasonably good agreement with the power law of 3/2.

$$\omega_+(\mathbf{k}) = \omega_c + \frac{\omega_T^2 + \omega_L^2}{2\omega_c} = \omega_c + \frac{1}{2}\left(\frac{e^2}{\hbar l_B}\right)\left(\frac{\nu\sqrt{3}}{\pi}\right)^{\frac{3}{2}} \mathrm{tr}\left[\widetilde{D}(\mathbf{k})\right],$$

showing the magnetophonon remains gapless while the magntoplasmon is gapped by $\omega_c$. Here, $\widetilde{D}(\mathbf{k}) = (ml_s^3/8e^2)\widehat{D}(\mathbf{k})$ is a dimensionless dynamical matrix whose magnitude is an order of unity [14].

The energy scaling laws of both of $\omega_{T\mathbf{k}}$ and $\omega_{L\mathbf{k}}$ in density tuning ($n \sim \nu B \sim 1/l_s^2$) are $\omega_0 \sim (8e^2/ml_s^3)^{1/2}$; thus, the magnetophonon is scaled as $\omega_- \sim 1/l_s^3 B \sim n^{3/2}/B \sim \sqrt{B}\,\nu^{3/2}$. This power law is the distinctive feature of the magnetophonon spectrum.

Probably the most remarkable feature of the scaling law of the magnetophonon is its carrier mass independence[11] (see Fig. S8). In the large field limit, the Lorentz force dominates the mass inertia, and the motion is defined by the balance between the Lorentz force and the Coulombic restoring force; the mass term in the equation of motion, Eq. (7), becomes negligible. In Fig. S8, we compare the spectra of electron and hole samples to find that the magnetophonon energy scale is comparable in magnitude even though the effective mass of holes is ~ 6 times heavier than that of electrons.

## 7. Analysis of the "vertical features" in tunneling current $I_t$ and $dI/dV$

The seemingly vertical features in data (see arrows in Fig. S11c and also in Fig. 1f) are consequences of an abrupt disappearance of the modified tunneling matrix element by electron-phonon coupling at the phase boundary. The increased noise suggests the existence of small hysteresis due to the first-order phase transition.

The tunneling current $I$ measured at bias voltage $V_t$ has the contributions from all electrons tunneling at energies between $eV_t$ and the Fermi level ($eV_t = 0$) (see Fig. S10a). In other words, the tunneling current $I$ is the summation of $dI/dV_t$ up to the bias energy $E_t = eV_t$, and given as (at $T = 0$)

$$I = \int_0^{eV_t} d(eV_t)\,\frac{dI}{edV_t} = \int_0^{eV_t} dE_t \int_{-\infty}^{\infty} d^2k\,\left|T(\vec{k}, E_F + E_t)\right|^2 \times A(\vec{k}, E_F + E_t),$$

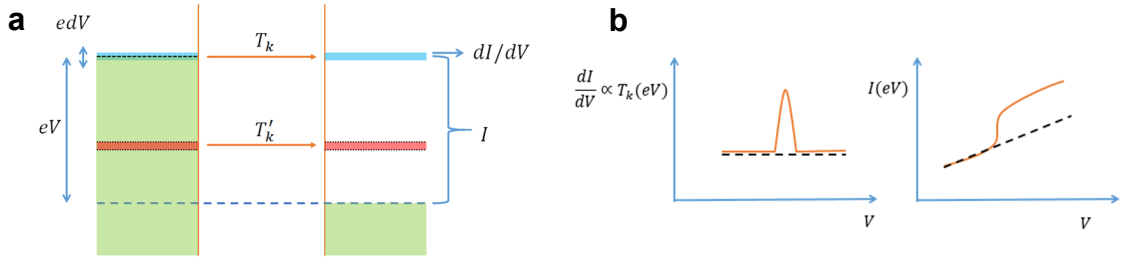

**Fig. S10 | Effect of tunneling matrix element change. a,** Tunneling current with a modified tunneling matrix element at energy $E_t = \epsilon_r$. **b,** Tunneling conductance and tunneling current with/without electron-boson coupling (orange/dotted). While the effect of a peak (or a dip) in the tunneling matrix only affects $dI/dV$ in the vicinity of the peak, the total tunneling current for injection voltages beyond the peak is kept increased from the bare value. If the peak suddenly disappears as the filling factor changes, it induces spurious vertical noise in otherwise smooth $dI/dV$ as shown by arrows in Fig. S11**c**.

and it is apparent that a change in $dI/dV_t$ (or $T(k,E)$) at certain tunneling bias energy $E_t = \epsilon_r$ should affect the measured current $I$ at subsequently higher bias voltages (see Fig. S10**b**). On the other hand, this is supposed to affect the differential conductance $dI/dV_t$ only around $E_t = \epsilon_r$. However, when there is an abrupt change in a component of the tunneling matrix while varying filling factor $\nu$, a big change in $I\ (E_t > \epsilon_{\nu H})$ can induce large noise in $dI/dV_t$, extending vertically at around the filling factor the tunneling matrix change occurs (See arrows in Fig. S11**c**).

Moreover, the existence of hysteresis in changing filling factor $\nu$ seems essential in having this noise pronounced. Even though the filling factor $\nu$ is strictly fixed along the vertical axis in the data presented, such as Fig. S11**c**, due to the nature of our feed-back based measurement scheme, there are intermediate measurement steps where the filling factor deviates from a desired value by some degree. If a small hysteresis exists, near the phase boundary, the intermediate feed-back steps will make the system end up randomly in either phase (either liquid or crystal) through uncontrolled phase transitions. This will induce large noise, again situated vertically (at around certain value of $\nu$) in the 2D image data of $dI/dV_t$ as shown in Fig. S11**b** and **c**. Note that without hysteresis, regardless of the intermediate steps, the phase of the system is well decided by the values $eV_t$ and $\nu$ of the final measurement step, and the noise will disappear. Note that in the limit of large hysteresis this noise will also disappear because the phase of the electron system is effectively fixed. Thus, $dI/dV_t$ image data plotted in $(eV_t, \nu)$ with vertical lines of increased noise is a consequence of a "small-size" hysteresis belonging to a first-order phase transition at that filling factor.

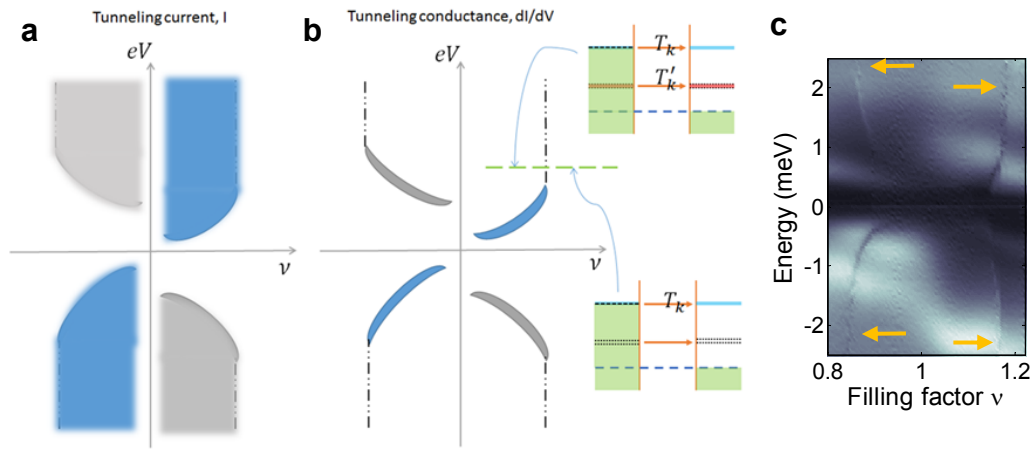

**Fig. S11 | Origin of the vertical features: First order phase transition. a-b,** Diagrams showing $I$ and $dI/dV$ in the case of sudden disappearances of phonons at the phase transition. TDOS (or $dI/dV$) can show spurious noise due to a sudden change in tunneling matrix $T_k$ (and possible hysteresis) at tunneling bias $eV$ near Fermi level. The noise will be greatly amplified with the existence of hysteresis in the phase transition. The existence of ${T'_k}'$ whose value is suppressed (or enhanced) from $T_k$ will determine the regions separated by the dotted vertical line in **b**.

## 8. Additional data plots (Figs. 12-14)

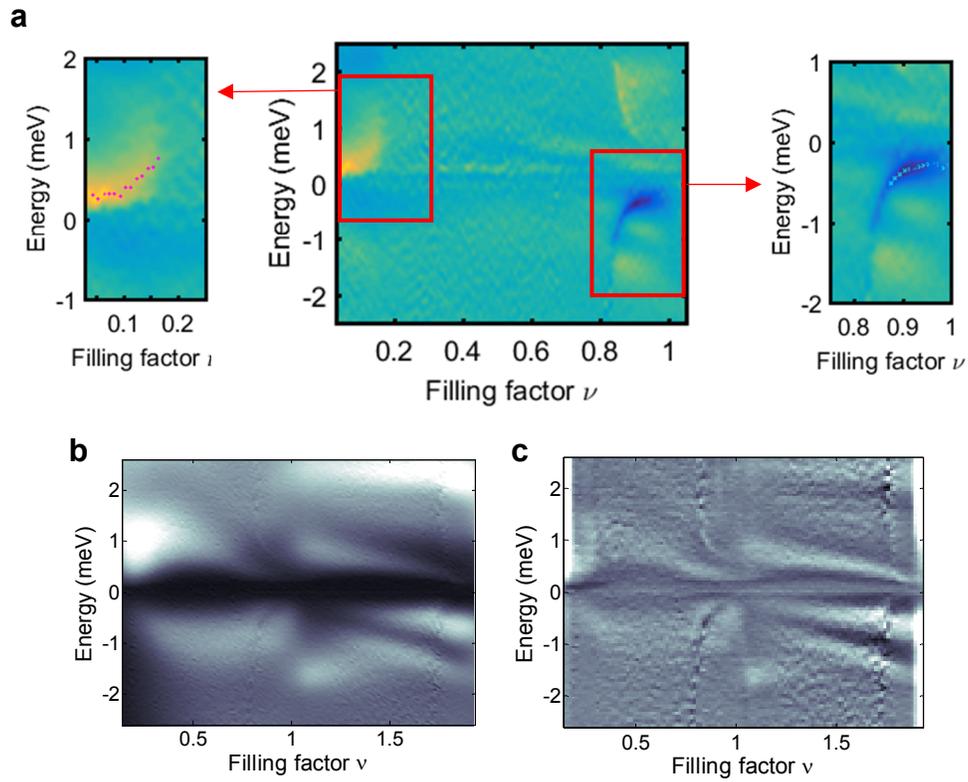

**Fig. S12 | Resonance features at even filling factors a,** Normalized TDOS of a hole system at 8 T. **b,** TDOS and **c,** Background-subtracted TDOS of a hole system at 5 T at T=25 mK.

(Continues on next page)

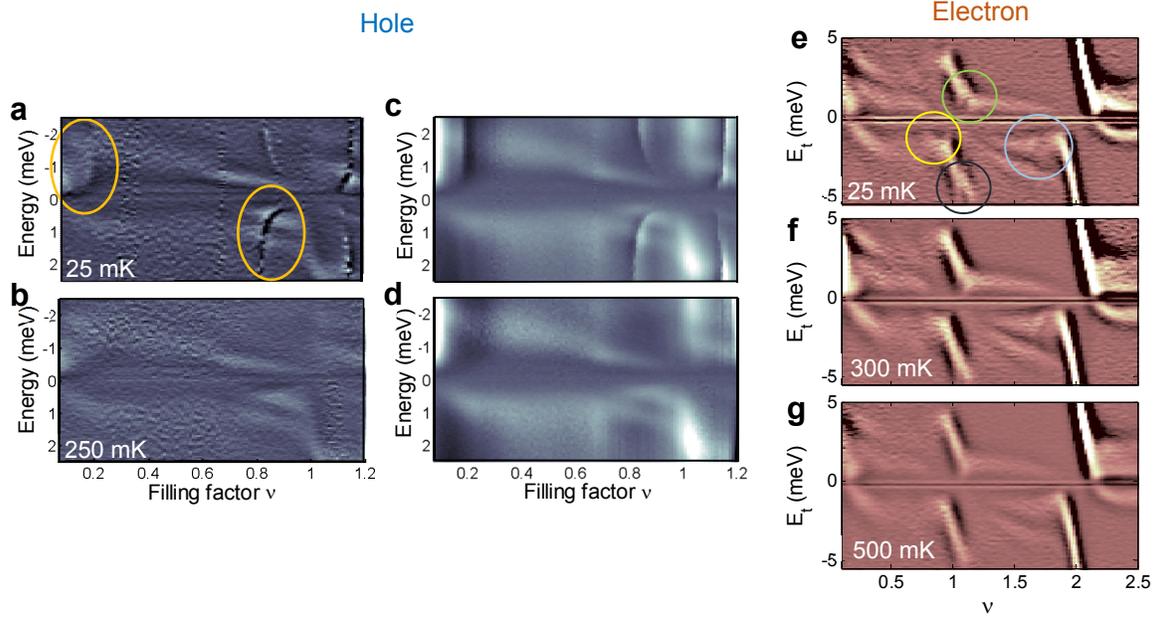

**Fig. S13 | Temperature dependence of the resonances. a-b,** Background subtracted conductance $dI/dV_t$ for the hole doped sample at $B_\perp = 8\,T$ at 25 mK and 250 mK. **c-d,** Background subtracted current $I$. The yellow circles in **a** emphasize that the energetics of the phonon features near $\nu \sim 0$ is particle-hole symmetric to ones near $\nu \sim 1$. **e-g,** TDOS of 2D electrons with $B_\perp = 5.8\,T, B_\parallel = 8.2\,T$ measured at 25 mK (**e**), 300 mK (**f**) and 500 mK (**g**). Spectra at **e**, **f** and **g** are after being subtracted with slowly-varying backgrounds to reveal fine structures of the spectrum. Raw data are shown in Fig. S14b. The colored circles highlight features of magnetophonon disappearing with increasing temperature.

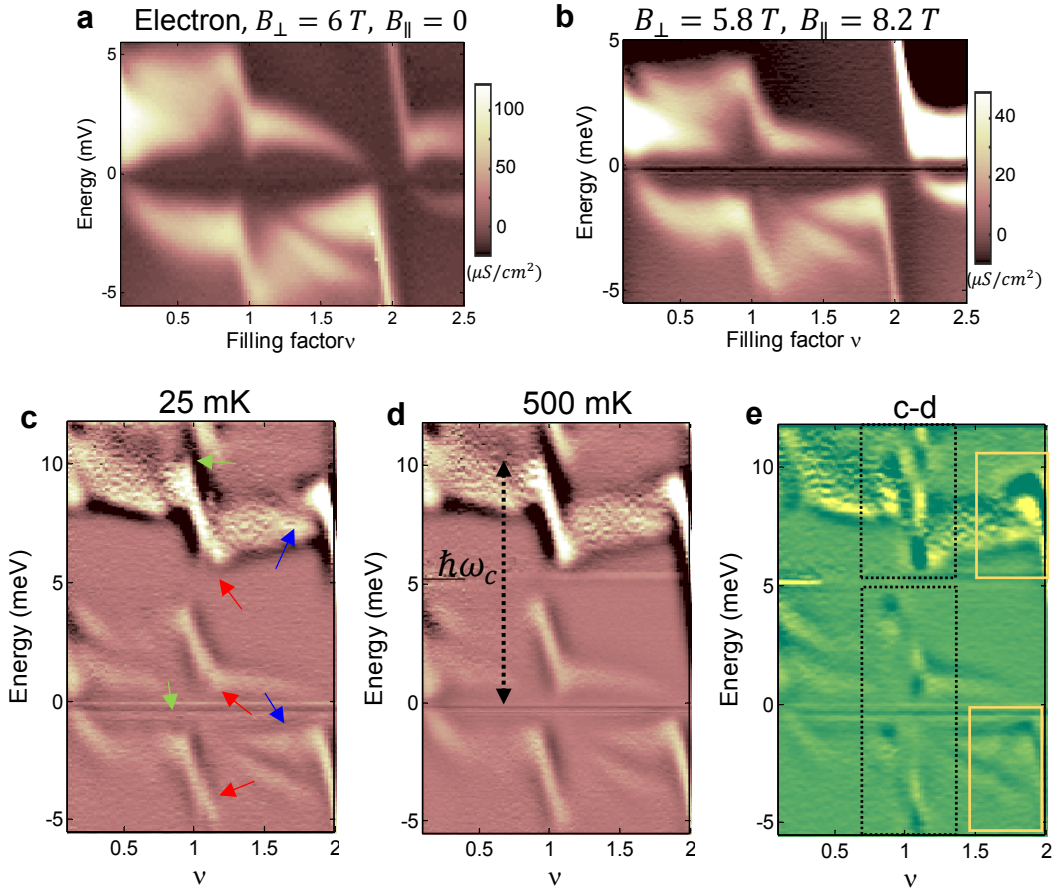

**Fig. S14 | The $dI/dV_t$ spectrum of electron-doped sample.** **a-b**, The raw spectra of Fig. 4 are plotted before being subtracted with a slowly-varying background, with $B_\perp = 6\,T, B_\parallel = 0\,T$ (**a**), and $B_\perp = 5.8\,T, B_\parallel = 8.2\,T$ (**b**). **c-e**, Background subtracted TDOS of 2D electrons with $B_\perp = 5.8\,T, B_\parallel = 8.2\,T$ is measured up to second Landau level at 25 mK (**c**), and 500 mK (**d**). The colored arrows highlight features existing only at low temperatures. In **e**, the spectrum in **c** is subtracted with the one in **d** to show strongly temperature dependent features shown inside squares.